\documentclass[letterpaper, 11pt]{article}
\usepackage[margin = 1in]{geometry}
\usepackage{epsfig,graphicx,color}
\usepackage[cp850]{inputenc}
\usepackage[T1]{fontenc}
\usepackage[english]{babel}
\usepackage{cite}
\usepackage{rotating}
\usepackage[f]{esvect}
\usepackage{amsmath, amsthm, amssymb,amsfonts}
\usepackage[f]{esvect}
\usepackage{lscape}
\usepackage{mathrsfs}
\usepackage{url}

\usepackage{mathrsfs}
\usepackage{commath}
\usepackage{pgfplots}
\usepackage{tikz}
\usepackage{rotfloat}
\usepackage{threeparttable}
\usepackage{xfrac}
\usepackage[tableposition=top]{caption}

\def\tablenotes{\bgroup\parfillskip=0pt plus 1fil
\leftskip=0pt\relax \rightskip=0pt
\vskip2pt\footnotesize}
\def\endtablenotes{\vskip1pt\egroup}

\def\sphline{\noalign{\vskip3pt}\hline\noalign{\vskip3pt}}

\theoremstyle{definition}

\begin{document}
\title{Compact Formulae for Three-Center Nuclear Attraction Integrals Over Exponential Type Functions}
\author{Richard M. Slevinsky$^{\S}$ and Hassan Safouhi$^\dag$\footnote{The corresponding author (HS) acknowledges the financial support from the Natural Sciences and Engineering Research Council of Canada~(NSERC) - Grant RGPIN-2016-04317}\\
\\
$^{\S}${\it Department of Mathematics, University of Manitoba}\\
{\it 186 Dysart Rd, Winnipeg (MB), Canada R3T 2M8}\\
{\it Richard.Slevinsky@umanitoba.ca}\\
\\
$^\dag${\it Campus Saint-Jean, University of Alberta}\\
{\it 8406, 91 Street, Edmonton (AB), Canada T6C 4G9}\\
{\it hsafouhi@ualberta.ca}
}

\date{ }

\maketitle

{\bf Abstract.} \hskip 0.15cm In any ab initio molecular orbital (MO) calculations, the major task involves the computation of the so-called molecular multi-center integrals. Multi-center integral calculations is a very challenging mathematical problem in nature. Quantum mechanics only determines which integrals we evaluate, but the techniques employed for their evaluations are entirely mathematical. The three-center nuclear attraction integrals occur in a very large number even for small molecules and are among of the most difficult molecular integrals to compute efficiently.

In the present contribution, we report analytical expressions for the three-center nuclear attraction integrals over exponential type functions. We describe how to compute the formula to obtain an efficient evaluation in double precision arithmetic. This requires the rational minimax approximants that minimize the maximum error on the interval of evaluation.

\vspace*{0.75cm}
{\bf Keywords:}

Three-center nuclear attraction integrals. Molecular integrals. Exponential type functions. Bessel functions. Molecular structure calculations.

\clearpage
\section{Introduction}

Molecular electronic structure theory is a highly interdisciplinary research topic whose progress depends crucially on the progress in numerical analysis, mainly numerical integration, and only a very few mathematicians were interested in this challenging mathematical problem. In this regards, Sack has played a prominent role with his work in the 1960s on addition theorems~\cite{Sack-5-252-64, Sack-5-260-64, Sack-8-1774-67, Sack-5-774-74}. Hellmann's book~\cite{Hellmann-1937} has a section dealing with the evaluation of these extremely difficult integrals. The pursuit of accurate and efficient algorithms for the numerical evaluation of molecular integrals for the purpose of electronic structure calculations has led to a substantial body of work. While molecular integrals with Gaussian type functions (GTFs)~\cite{Boys-200-542-50, Boys-201-125-50} as a basis have been for a long time the most easily calculated, their theoretical deficiency both in the long and short ranges~\cite{Agmon-85, Kato-10-151-57} has led to a renewed interest in an exponential type function (ETF) basis. ETFs are better suited than GTFs to represent electron wave functions near the nucleus and at long range, provided that multi-center integrals using such functions could be computed efficiently.
Examples of such a basis include the Slater type functions (STFs)~\cite{Slater-42-33-32}, the Laguerre type functions, and the $B$ functions~\cite{Shavitt-63, Steinborn-Filter-38-273-75, Filter-Steinborn-18-1-78}. We note that many researchers hope that the next generation of ab initio programs will be based on the usage of ETFs. Indeed much effort is being made to develop efficient molecular algorithms for integrals over conventional ETFs~(STFs or $B$ functions)~\cite{Ozdogan-Ruiz-08, Niehaus-Lopez-Rico-41-485205-08, Fernandez-Ema-Lopez-Ramirez-78-83-00, Rico-Fernandez-Ema-Lopez-Ramirez-81-16-01, Fernandez-Aguado-Lopez-Ema-Ramirez-8-148-01, Steinborn-Homeier-Ema-Lopez-Ramirez-76-244-00} and references therein.

The $B$ function basis has been at the forefront of recent developments~\cite{Steinborn-Weniger-11-509-77, Steinborn-Weniger-12-103-78, Weniger-Steinborn-78-6121-83, Weniger-Grotendorst-Steinborn-33-3688-86, Grotendorst-Weniger-Steinborn-33-3706-86, Weniger-Steinborn-73-323-88}, as their simple Fourier transform leads to compact analytical expressions for the integrals. Unfortunately, through the expansions, some of the resulting semi-infinite integrals have integrands composed of the Bessel functions. As the most complicated part of the analytical expressions, the semi-infinite integrals have become the bottleneck of the calculations in the $B$ function approach. Since the identification of this computational problem, a large body of work is devoted to dealing with this bottleneck. The most common element to most algorithms~\cite{Safouhi9, Safouhi21, Safouhi31, Safouhi32, Safouhi35, Safouhi38} involves an intense integration-then-extrapolation approach, which is usually characterized by subdividing the integral between the oscillatory Bessel function's zeros, by integrating by a quadrature, and using the resulting sequence to estimate the remainder of the integral. Of course, many improvements to the general extrapolation procedure are documented and indeed some of the most recent of these improvements are considered state-of-the-art for multi-center integrals. While these methods and their refinements are generally highly accurate and efficient, there are some ranges of parameters where either failure is inevitable or the computation becomes extremely heavy.

In the comparative study~\cite{Safouhi45} of the most popular extrapolation methods and sequence transformations for computing semi-infinite integrals, the authors conclude that having asymptotic series representations for integrals and applying sequence transformations to accelerate their convergence or to sum their divergence leads to the most efficient algorithms for computing the integrals. However, when such asymptotic series representations do not exist, refinements to either the numerical steepest descent method~\cite{Iserles-Norsett-44-755-04, Iserles-Norsett-461-1383-05, Olver-26-213-06, Huybrechs-Vandewalle-44-1026-06, Cools-Huybrechs-Nuyens-109-1748-09, Huybrechs-Olver-09} or the extrapolation methods~\cite{Levin-Sidi-9-175-81, Gray-Atchison-4-363-67, Brezinski-Zaglia-91, Gray-Wang-29-271-92,Sidi-03} must be made to obtain a desirable outcome. This conclusion does not particularly challenge any preconceived notions. However, it does emphasize that the ultimate goal in computing semi-infinite integrals is to find analytical expressions. Several examples of this approach have been documented in the literature on molecular integrals in the $B$ function basis~\cite{Weniger-Steinborn-78-6121-83, Safouhi39}. The expressions that are obtained have greatly simplified the calculation of one- and two-center integrals. The pursuit of such analytical expressions stopped in the 1990s at the three-center integrals because of the increased complexity of the integrands.

In this work, we report a significant breakthrough for the three-center nuclear attraction integrals. This breakthrough takes the form of an analytical expression for the semi-infinite integrals. In section 2, we introduce the general definitions and properties required for the discussion. In section 3, we prove our main result. In section 4, we hold a numerical discussion on its performance and compare it to existing methods based on extrapolation. Our comparison leads us to believe that the analytical expression is useful in a numerical setting and performs extremely well. Indeed, the decrease in calculation time is substantial compared with the existing methods.

Numerical experiments conducted for the integrals in~\eqref{eq:TCcyl}, show that the use of the analytic expressions is $10^{3}$ times more efficient than the state-of-the-art~[S7].

Since the nuclear magnetic resonance (NMR) integrals, such as first order integrals in the relativistic calculations of the nuclear shielding tensor~\cite{Safouhi33, Safouhi41}, are also expressed in terms of semi-infinite integrals similar to three-center nuclear attraction integrals, the proposed analytic development will also have a major impact in computing NMR properties of molecular systems. In addition, on the algorithmic side, existing density functional theory based programs, such as ADF (Amsterdam Density Functional) will benefit from the results of this research, since even in such an approach, integrals similar to three-center nuclear attraction are needed. The proposed methods could also be implemented in the existing frameworks, such as Slater type orbital package~(STOP) and SMILE (Slater molecular integrals for large electronic systems) where STFs are expanded in terms of GTFs.

\section{General definitions and properties}
The functions $B_{n,l}^m(\zeta,\vv{r})$ are defined by~\cite{Steinborn-Filter-38-273-75, Filter-Steinborn-18-1-78}:
\begin{equation}
B_{n,l}^m(\zeta,\vv{r}) = \frac{(\zeta r)^l}{2^{n+l}(n+l)!} \, \hat{k}_{n-\frac{1}{2}}(\zeta r) \, Y_l^m(\theta_{\vv{r}},\varphi_{\vv{r}}),
\end{equation}
where $n,l,$ and $m$ are the quantum numbers such that $n=1,2,\ldots,$ $l=0,1,\ldots,n-1,$ and $m=-l,-l+1,\ldots l-1,l$ and where $Y_l^m(\theta,\phi)$ denotes the surface spherical harmonic, and is defined explicitly using the Condon-Shortley phase convention as~\cite{Condon-Shortley-51}:
\begin{equation}
Y_l^m(\theta,\phi) = {\rm i}^{m+|m|}\sqrt{\dfrac{(2l+1)(l-|m|)!}{4\pi(l+|m|)!}}P_l^{|m|}(\cos\theta)e^{{\rm i}m\phi},
\end{equation}
where $P_l^m(x)$ is the associated Legendre polynomial of $l^{\rm th}$ degree and $m^{\rm th}$ order:
\begin{equation}
P_l^m(x) = (1-x^2)^{m/2}\dfrac{{\rm d}^{l+m}}{{\rm d}x^{l+m}}\left(\dfrac{(x^2-1)^l}{2^l\,l!}\right).
\end{equation}

The radial part of the $B$ functions is given by the reduced Bessel functions $\hat{k}_{n-\frac{1}{2}}(\zeta r)$. For half-integral orders, the reduced Bessel functions can be represented by an exponential multiplied by a terminating confluent hypergeometric function~\cite{Steinborn-Filter-38-273-75, Filter-Steinborn-18-1-78}:
\begin{equation}
\hat{k}_{n+\frac{1}{2}}(z) = 2^n(1/2)_n e^{-z} {}_1F_1(-n;-2n;2z),
\end{equation}
where $(x)_n$ stands for the Pochhammer symbol, which may be defined in terms of the Gamma function $\Gamma(z)$ as~\cite{Abramowitz-Stegun-65}:
\begin{equation}
(x)_n = \dfrac{\Gamma(x+n)}{\Gamma(x)}.
\end{equation}

Reduced Bessel functions are related to modified Bessel functions of the second kind by~\cite{Watson-44}:
\begin{equation}
\hat{k}_\nu(z) = \sqrt{\frac{2}{\pi}}z^{\nu}K_{\nu}(z),
\end{equation}
Reduced and modified Bessel functions satisfy the following three-term recurrence relations~\cite{Abramowitz-Stegun-65}:
\begin{align}
\hat{k}_{\nu+1}(z) & = 2\nu\hat{k}_\nu(z) + z^2 \hat{k}_{\nu-1}(z) \qquad \textrm{and} \qquad
K_{\nu+1}(z) = \dfrac{2\nu}{z}K_\nu(z) + K_{\nu-1}(z).
\end{align}

The integral representation of the modified Bessel function is given by~\cite{Abramowitz-Stegun-65}:
\begin{equation}
\dfrac{K_\nu(xz)}{z^\nu}  = \dfrac{1}{2}\int_0^\infty\dfrac{e^{-\frac{x}{2}(t+\frac{z^2}{t})}}{t^{\nu+1}}{\rm\,d}t
\label{EQINTREPMODBESS}
\end{equation}

The modified Bessel functions of the second kind satisfy the following property for some $m\in\mathbb{N}_0$~\cite{Abramowitz-Stegun-65}:
\begin{align}
\left(\dfrac{\rm d}{z{\rm\,d}z}\right)^m\left(z^{\pm\nu} K_\nu(z)\right) & = (-1)^mz^{\pm\nu-m}K_{\nu\mp m}(z).
\label{eq:Krecrel}
\end{align}

The normalized STFs are defined by~\cite{Slater-42-33-32}:
\begin{equation}
\chi_{n,l}^{m}(\zeta,\vv{r})\;=\; {\cal N}(\zeta,n) \, r^{n-1} \, e^{-\zeta r} \,
Y_{l}^{m}(\theta_{\vv{r}},\varphi_{\vv{r}}),
\label{EQSLAT}
\end{equation}
where $N(\zeta,n)$ stands for the normalization factor and it is given by:
\begin{equation}
{\cal N}(\zeta,n) \,=\, \sqrt{\frac{(2\zeta)^{2n+1}}{(2n)!}}.
\end{equation}
STFs can be expressed as finite linear combinations of $B$ functions \cite{Filter-Steinborn-18-1-78}:
\begin{equation}
\chi_{n,l}^{m}(\zeta,\vv{r})\;=\; \frac{{\cal N}(\zeta,n)}{\zeta^{n-1}} \, \sum_{p=\tilde{p}}^{n-l} \,
\frac{(-1)^{n-l-p} \,\, 2^{2 p + 2 l - n} \,\, (l+p)!}{(2p-n+l)! \,\, (n-l-p)!}\,
B_{p,l}^{m}(\zeta,\vv{r}),
\label{EQSLBF}
\end{equation}
where:
\begin{equation}
\tilde p =
\left\lbrace
\begin{array}{llll}
\displaystyle \frac{n-l}{2}   \quad & \textrm{if} & \quad n-l \quad  & \textrm{is even}\\
\displaystyle \frac{n-l+1}{2} \quad & \textrm{if} & \quad n-l \quad  & \textrm{is odd}.
\end{array}
\right.
\label{EQPTILDE}
\end{equation}

A given function $f(\vec{r})$ and its Fourier transform $\bar{f}(\vec{k})$ are connected by the symmetric relationships~\cite{Arfken-Weber-95}:
\begin{equation}
\bar{f}(\vec{k}) = (2\pi)^{-3/2} \int_{\vec{r}} e^{-i \vec{k} \cdot \vec{r}} \; f(\vec{r}) \; d\vec{r} \qquad \textrm{and} \qquad f(\vec{r}) = (2\pi)^{-3/2} \int_{\vec{k}} e^{~i \vec{k} \cdot \vec{r}} \; \bar{f}(\vec{k}) \; d\vec{k}.
\label{eq:TFDef}
\end{equation}

Consequently, the Rayleigh expansion of the plane wavefunctions is useful~\cite{Weissbluth-78}:
\begin{equation}
e^{\pm{\rm i}\vec{p}\cdot\vec{r}} = \sum_{l=0}^\infty\sum_{m=-l}^l 4\pi(\pm{\rm i})^l j_l(|\vec{p}||\vec{r}|)Y_l^m(\theta_{\vec{r}},\phi_{\vec{r}})[Y_l^m(\theta_{\vec{p}},\phi_{\vec{p}})]^*,
\end{equation}
where the spherical Bessel function $j_n(x)$ of order $n\in\mathbb{N}_0$ is defined by~\cite{Abramowitz-Stegun-65}:
\begin{equation}
j_n(x) = (-x)^n\left(\dfrac{{\rm d}}{x{\rm\,d}x}\right)^n\left(\dfrac{\sin(x)}{x}\right).
\end{equation}

Spherical Bessel functions are related to Bessel functions of the first kind $J_{n+1/2}(x)$ by~\cite{Abramowitz-Stegun-65}:
\begin{equation}
j_n(x) = \sqrt{\frac{\pi}{2\,x}}J_{n+1/2}(x).
\end{equation}

Spherical Bessel functions and Bessel functions of the first kind satisfy the three-term recurrence relations~\cite{Abramowitz-Stegun-65}:
\begin{equation}
j_{n+1}(x) = \dfrac{2n+1}{x} j_n(x) - j_{n-1}(x) \qquad \textrm{and} \qquad J_{\eta+1}(x) = \dfrac{2\eta}{x}J_\eta(x) - J_{\eta-1}(x).
\end{equation}

The Bessel functions satisfy the following properties~\cite{Abramowitz-Stegun-65}:
\begin{align}
J_\eta(z) & = \sum_{k=0}^\infty\dfrac{(-1)^k}{k!\,\Gamma(k+\eta+1)}\left(\dfrac{z}{2}\right)^{2k+\eta}
\nonumber\\
\left(\dfrac{\rm d}{z{\rm\,d}z}\right)^m\left(z^\eta J_\eta(z)\right) & = z^{\eta-m}J_{\eta-m}(z) \qquad \textrm{for} \qquad m\in\mathbb{N}_0.
\label{eq:Jrecrel}
\end{align}

Due to the Rayleigh expansion, the $B$ functions have a relatively simple Fourier transform~\cite{Weniger-Steinborn-78-6121-83}:
\begin{equation}
\bar{B}_{n,l}^m(\zeta,\vec{p}) = \sqrt{\dfrac{2}{\pi}}\zeta^{2n+l-1}\dfrac{(-{\rm i}p)^l}{(\zeta^2+p^2)^{n+l+1}}Y_l^m(\theta_{\vec{p}},\phi_{\vec{p}}).
\end{equation}

The Fourier integral representation of the Coulomb operator is given by~\cite{Gelfand-Shilov-64}:
\begin{equation}
\dfrac{1}{|\vec{r}-\vec{R}_1|} = \dfrac{1}{2\pi^2}\int_{\vec{p}} \dfrac{e^{-{\rm i}\vec{p}\cdot(\vec{r}-\vec{R}_1)}}{p^2}{\rm\,d}^3\vec{p}.
\end{equation}

The Gaunt coefficients are defined as~\cite{Gaunt-228-151-29,Weniger-Steinborn-25-149-82,Xu-65-1601-96,Xu-85-53-97}:
\begin{equation}
\langle l_1m_1|l_2m_2|l_3m_3\rangle = \int_0^{2\pi}\int_0^\pi [Y_{l_1}^{m_1}(\theta,\phi)]^*Y_{l_2}^{m_2}(\theta,\phi)Y_{l_3}^{m_3}(\theta,\phi) \sin\theta{\rm\,d}\theta{\rm\,d}\phi.
\end{equation}

These coefficients linearize the product of two spherical harmonics:
\begin{equation}
[Y_{l_1}^{m_1}(\theta,\phi)]^*Y_{l_2}^{m_2}(\theta,\phi) = \sum_{l=l_{\rm min},2}^{l_1+l_2} \langle l_2m_2|l_1m_1|lm_2-m_1\rangle Y_l^{m_2-m_1}(\theta,\phi),
\end{equation}
where the subscript $l=l_{\rm min},2$ in the summation implies that the summation index $l$ runs in steps of $2$ from $l_{\rm min}$ to $l_1+l_2$, and the constant $l_{\rm min}$ is given by:
\begin{equation}\label{EQSUMINDEX}
l_{\rm min} = \left\{\begin{array}{lccr}
\max(|l_1-l_2|,|m_2-m_1|)    &{\rm if}& l_1+l_2+\max(|l_1-l_2|,|m_2-m_1|)&{\rm is~even}\\
\max(|l_1-l_2|,|m_2-m_1|)+1  &{\rm if}& l_1+l_2+\max(|l_1-l_2|,|m_2-m_1|)&{\rm is~odd.}\\
\end{array}\right.
\end{equation}

The three-center nuclear attraction integrals over $B$ functions are given by:
\begin{equation}
{\cal I}_{n_1,l_1,m_1}^{n_2,l_2,m_2} = \int_{\vec{R}}\left[B_{n_1,l_1}^{m_1}(\zeta_1,\vec{R}-\overrightarrow{OA})\right]^*\dfrac{1}{|\vec{R}-\overrightarrow{OC}|}B_{n_2,l_2}^{m_2}(\zeta_2,\vec{R}-\overrightarrow{OB}){\rm\,d}^3\vec{R},
\end{equation}
where $A$, $B$, and $C$ are three arbitrary points of $\mathbb{R}^3$, and $O$ is the origin. By performing a translation of $\overrightarrow{OA}$ and by substituting the integral representation of the Coulomb operator in the above equation, the integrals can be re-written as:
\begin{equation}
{\cal I}_{n_1,l_1,m_1}^{n_2,l_2,m_2} = \dfrac{1}{2\pi^2}\int_{\vec{x}}\dfrac{e^{{\rm i}\vec{x}\cdot\vec{R}_1}}{x^2}\left\langle B_{n_1,l_1}^{m_1}(\zeta_1,\vec{r})\left|e^{-{\rm i}\vec{x}\cdot\vec{r}}\right|B_{n_2,l_2}^{m_2}(\zeta_2,\vec{r}-\vec{R}_2)\right\rangle_{\vec{r}}{\rm\,d}^3\vec{x},
\end{equation}
where:
\begin{equation}
\left\langle B_{n_1,l_1}^{m_1}(\zeta_1,\vec{r})\left|e^{-{\rm i}\vec{x}\cdot\vec{r}}\right|B_{n_2,l_2}^{m_2}(\zeta_2,\vec{r}-\vec{R}_2)\right\rangle_{\vec{r}} = \int_{\vec{r}}\left[B_{n_1,l_1}^{m_1}(\zeta_1,\vec{r})\right]^*e^{-{\rm i}\vec{x}\cdot\vec{r}} B_{n_2,l_2}^{m_2}(\zeta_2,\vec{r}-\vec{R}_2){\rm\,d}^3\vec{r},
\end{equation}
and where $\vec{r} = \vec{R}-\overrightarrow{OA}$, $\vec{R}_1 = \overrightarrow{OC}$, and $\vec{R}_2 = \overrightarrow{AB}$.

Using all of the above properties, the three-center nuclear attraction integrals over $B$ functions have the representation~\cite{Trivedi-Steinborn-27-670-83,Grotendorst-Steinborn-38-3875-88}:
\begin{eqnarray}
{\cal I}_{n_1,l_1,m_1}^{n_2,l_2,m_2} & = & \frac{8 \, (4 \, \pi)^2 \, (-1)^{l_1+l_2} \, (2l_1+1)!! \, (2l_2+1)!! \, (n_1+l_1+n_2+l_2+1)! \, \zeta_1^{2n_1+l_1-1} \, \zeta_2^{2n_2+l_2-1}}{(n_1+l_1)!(n_2+l_2)!}
\nonumber\\ & \times & \sum_{l_1^{\prime}=0}^{l_1} \sum_{m_1^{\prime}=-l_1^{\prime}}^{l_1^{\prime}} ({\rm i})^{l_1+l_1^{\prime}}\, \frac{\langle l_1m_1|l_1^{\prime}m_1^{\prime}|l_1-l_1^{\prime}m_1-m_1^{\prime}\rangle}{(2\,l_1^{\prime}+1)!! \, [2\, (l_1-l_1^{\prime})+1]!!}
\nonumber\\ & \times & \sum_{l_2^{\prime}=0}^{l_2} \sum_{m_2^{\prime}=-l_2^{\prime}}^{l_2^{\prime}} ({\rm i})^{l_2+l_2^{\prime}}\, (-1)^{l_2^{\prime}}
\frac{\langle l_2m_2|l_2^{\prime}m_2^{\prime}|l_2-l_2^{\prime}m_2-m_2^{\prime}\rangle}{(2\, l_2^{\prime}+1)!!\, [2\, (l_2-l_2^{\prime})+1]!!}
\nonumber\\ & \times & \sum_{l=l_{\min}^{\prime},2}^{l_{2}^{\prime}+l_{1}^{\prime}}
\langle l_2^{\prime} m_2^{\prime}|l_1^{\prime}m_1^{\prime}|lm_2^{\prime}-m_1^{\prime}\rangle\,
R_2^l \, Y_{l}^{m_2^{\prime}-m_1^{\prime}}(\theta_{\vec{R}_2},\varphi_{\vec{R}_2})
\nonumber\\ & \times & \sum_{\lambda=l_{\min}^{\prime \prime},2}^{l_{2}-l_{2}^{\prime}+l_{1}-l_{1}^{\prime}} (-{\rm i})^{\lambda}
\langle l_2-l_2^{\prime}m_2-m_2^{\prime}|l_1-l_1^{\prime}m_1-m_1^{\prime}|\lambda \, \mu\rangle
\nonumber\\ & \times & \sum_{j=0}^{\Delta l}\, {\Delta l \choose j}\, \frac{(-1)^{j}}{2^{n_1+n_2+l_1+l_2-j+1} \, (n_1+n_2+l_1+l_2-j+1)!}
\nonumber\\ & \times & \int_{0}^{1} s^{n_2+l_2+l_1-l_1^{\prime}} \, (1-s)^{n_1+l_1+l_2-l_2^{\prime}}
\, Y_{\lambda}^{\mu}(\theta_{\vec{v}},\varphi_{\vec{v}}) \,
\nonumber\\ &  & \hskip 1cm \times
\left[ \int_{0}^{+\infty} \, x^{n_x} \, \frac{\hat{k}_{\nu}[R_2 \gamma_{12}(s,x)]}{[\gamma_{12}(s,x)]^{n_{\gamma}}} \, j_{\lambda}(v\,x){\rm\,d}x \right] {\rm\,d}s,
\label{EQANAT}
\end{eqnarray}
where :
\begin{align}
[\gamma_{12}(s,x)]^2 & = (1-s)\zeta_{1}^{2}+s\zeta_{2}^{2}+s(1-s)x^{2} \nonumber\\
n_{\gamma} & = 2(n_1+l_1+n_2+l_2)-(l_1^{\prime}+l_2^{\prime})-l+1 \nonumber\\
\nu & = n_1+n_2+l_1+l_2-l-j+\frac{1}{2} \nonumber\\
\mu & = (m_2-m_2^{\prime})-(m_1-m_1^{\prime}) \nonumber\\
n_x & = l_1-l_1^{\prime}+l_2-l_2^{\prime} \nonumber\\
\Delta l & =[(l_1^{\prime}+l_2^{\prime}-l)/2] \nonumber\\
\vec{v} & = (1-s)\vec{R}_{2}-\vec{R}_{1},
\label{EQANATPARAM}
\end{align}
and where $v$ and $R_{2}$ stand for the modulus of $\vec{v}$ and $\vec{R}_{2}$ respectively.

The bottleneck in the numerical evaluation of this expression is the semi-infinite integral:
\begin{equation}
{\cal I}(s) = \int_{0}^{+\infty} \, x^{n_x} \, \frac{\hat{k}_{\nu}[R_2 \gamma_{12}(s,x)]}{[\gamma_{12}(s,x)]^{n_{\gamma}}} \, j_{\lambda}(v\,x){\rm\,d}x,
\label{EQANATSIBESS}
\end{equation}
which has varying degrees of oscillation and attenuation depending upon the values of the parameters.

The semi-infinite integrals ${\cal I}(s)$~\eqref{EQANATSIBESS} involves spherical Bessel functions $j_{\lambda}(v\,x)$ and this is why their accurate and rapid numerical evaluation emerge as the critical calculation in the analytical expressions, in particular for values of $s$ close to $0$ or $1$. If we make the substitution $s=0$ or $1$ in ${\cal I}_{s}$, $\gamma_{12}(s,x)$ becomes a constant and the integrand will be reduced to $x^{n_{x}} \, j_{\lambda}(v\,x)$,
because the exponential decreasing part $\hat{k}_{\nu}$ of the integrand becomes a constant and consequently the rapid oscillations of $j_{\lambda}(v \,x)$ cannot be damped and suppressed as it can be seen from Figure~\eqref{NucAttrac}. In addition, when the value of $v$ is large, the zeros of the integrands become closer and the oscillations become strong and the accurate numerical evaluation of the semi-infinite integrals becomes extremely challenging.

\begin{figure}[H]
	\centering
	\begin{tikzpicture}[scale=0.750]
	\begin{axis}[
	axis lines = left,
	xlabel={$x$},
	ylabel={Integrand of  $\mathcal{I}(s)$ ($10^{-3}$)},
	xmin=0, xmax=35.5,
	ymin=-150.5, ymax=150.5,
	ymajorgrids=true,
	grid style=dashed,
	]
	\addplot[no marks]
	coordinates {
( 0.000,      0.0000)  ( 0.143,      0.0000)  ( 0.286,      0.0000)  ( 0.429,      0.0000)  ( 0.572,      0.0001)  ( 0.715,      0.0006)
( 0.858,      0.0026)  ( 1.001,      0.0083)  ( 1.144,      0.0225)  ( 1.287,      0.0530)  ( 1.430,      0.1120)  ( 1.573,      0.2156)
( 1.716,      0.3836)  ( 1.859,      0.6369)  ( 2.002,      0.9936)  ( 2.145,      1.4641)  ( 2.288,      2.0446)  ( 2.431,      2.7112)
( 2.574,      3.4139)  ( 2.717,      4.0742)  ( 2.860,      4.5854)  ( 3.003,      4.8182)  ( 3.146,      4.6313)  ( 3.289,      3.8873)
( 3.432,      2.4725)  ( 3.575,      0.3198)  ( 3.718,     -2.5693)  ( 3.861,     -6.1060)  ( 4.004,    -10.1016)  ( 4.147,    -14.2660)
( 4.290,    -18.2170)  ( 4.433,    -21.5041)  ( 4.576,    -23.6448)  ( 4.719,    -24.1724)  ( 4.862,    -22.6908)  ( 5.005,    -18.9309)
( 5.148,    -12.8031)  ( 5.291,     -4.4384)  ( 5.434,      5.7885)  ( 5.577,     17.2586)  ( 5.720,     29.1340)  ( 5.863,     40.4106)
( 6.006,     49.9950)  ( 6.149,     56.8006)  ( 6.292,     59.8551)  ( 6.435,     58.4096)  ( 6.578,     52.0369)  ( 6.721,     40.7086)
( 6.864,     24.8419)  ( 7.007,      5.3054)  ( 7.150,    -16.6167)  ( 7.293,    -39.3049)  ( 7.436,    -60.9237)  ( 7.579,    -79.5710)
( 7.722,    -93.4462)  ( 7.865,   -101.0209)  ( 8.008,   -101.1967)  ( 8.151,    -93.4338)  ( 8.294,    -77.8359)  ( 8.437,    -55.1818)
( 8.580,    -26.8966)  ( 8.723,      5.0374)  ( 8.866,     38.2245)  ( 9.009,     70.0436)  ( 9.152,     97.8653)  ( 9.295,    119.2785)
( 9.438,    132.3065)  ( 9.581,    135.5935)  ( 9.724,    128.5413)  ( 9.867,    111.3851)  (10.010,     85.1978)  (10.153,     51.8219)
(10.296,     13.7328)  (10.439,    -26.1556)  (10.582,    -64.7266)  (10.725,    -98.9179)  (10.868,   -125.9777)  (11.011,   -143.6973)
(11.154,   -150.6000)  (11.297,   -146.0693)  (11.440,   -130.4044)  (11.583,   -104.7996)  (11.726,    -71.2473)  (11.869,    -32.3747)
(12.012,      8.7729)  (12.155,     48.9824)  (12.298,     85.1355)  (12.441,    114.4625)  (12.584,    134.7630)  (12.727,    144.5784)
(12.870,    143.2999)  (13.013,    131.2052)  (13.156,    109.4215)  (13.299,     79.8192)  (13.442,     44.8469)  (13.585,      7.3222)
(13.728,    -29.8023)  (13.871,    -63.6789)  (14.014,    -91.7897)  (14.157,   -112.1389)  (14.300,   -123.3967)  (14.443,   -124.9852)
(14.586,   -117.1006)  (14.729,   -100.6723)  (14.872,    -77.2636)  (15.015,    -48.9257)  (15.158,    -18.0170)  (15.301,     12.9953)
(15.444,     41.7344)  (15.587,     66.0977)  (15.730,     84.4117)  (15.873,     95.5448)  (16.016,     98.9731)  (16.159,     94.7937)
(16.302,     83.6883)  (16.445,     66.8423)  (16.588,     45.8265)  (16.731,     22.4557)  (16.874,     -1.3665)  (17.017,    -23.8011)
(17.160,    -43.2140)  (17.303,    -58.2933)  (17.446,    -68.1333)  (17.589,    -72.2834)  (17.732,    -70.7571)  (17.875,    -64.0041)
(18.018,    -52.8501)  (18.161,    -38.4100)  (18.304,    -21.9835)  (18.447,     -4.9429)  (18.590,     11.3778)  (18.733,     25.7849)
(18.876,     37.3080)  (19.019,     45.2614)  (19.162,     49.2783)  (19.305,     49.3188)  (19.448,     45.6502)  (19.591,     38.8063)
(19.734,     29.5273)  (19.877,     18.6887)  (20.020,      7.2236)  (20.163,     -3.9533)  (20.306,    -14.0150)  (20.449,    -22.2799)
(20.592,    -28.2540)  (20.735,    -31.6549)  (20.878,    -32.4182)  (21.021,    -30.6861)  (21.164,    -26.7808)  (21.307,    -21.1660)
(21.450,    -14.4002)  (21.593,     -7.0864)  (21.736,      0.1778)  (21.879,      6.8458)  (22.022,     12.4602)  (22.165,     16.6806)
(22.308,     19.3006)  (22.451,     20.2527)  (22.594,     19.6025)  (22.737,     17.5321)  (22.880,     14.3171)  (23.023,     10.2973)
(23.166,      5.8452)  (23.309,      1.3350)  (23.452,     -2.8868)  (23.595,     -6.5253)  (23.738,     -9.3554)  (23.881,    -11.2332)
(24.024,    -12.1000)  (24.167,    -11.9792)  (24.310,    -10.9675)  (24.453,     -9.2207)  (24.596,     -6.9371)  (24.739,     -4.3378)
(24.882,     -1.6482)  (25.025,      0.9198)  (25.168,      3.1828)  (25.311,      4.9975)  (25.454,      6.2676)  (25.597,      6.9475)
(25.740,      7.0405)  (25.883,      6.5944)  (26.026,      5.6936)  (26.169,      4.4490)  (26.312,      2.9875)  (26.455,      1.4403)
(26.598,     -0.0672)  (26.741,     -1.4247)  (26.884,     -2.5437)  (27.027,     -3.3626)  (27.170,     -3.8484)  (27.313,     -3.9967)
(27.456,     -3.8291)  (27.599,     -3.3892)  (27.742,     -2.7373)  (27.885,     -1.9434)  (28.028,     -1.0818)  (28.171,     -0.2245)
(28.314,      0.5641)  (28.457,      1.2310)  (28.600,      1.7379)  (28.743,      2.0626)  (28.886,      2.1988)  (29.029,      2.1553)
(29.172,      1.9536)  (29.315,      1.6250)  (29.458,      1.2075)  (29.601,      0.7417)  (29.744,      0.2679)  (29.887,     -0.1771)
(30.030,     -0.5627)  (30.173,     -0.8658)  (30.316,     -1.0719)  (30.459,     -1.1756)  (30.602,     -1.1797)  (30.745,     -1.0943)
(30.888,     -0.9352)  (31.031,     -0.7222)  (31.174,     -0.4772)  (31.317,     -0.2222)  (31.460,      0.0226)  (31.603,      0.2396)
(31.746,      0.4153)  (31.889,      0.5409)  (32.032,      0.6121)  (32.175,      0.6295)  (32.318,      0.5974)  (32.461,      0.5236)
(32.604,      0.4182)  (32.747,      0.2926)  (32.890,      0.1586)  (33.033,      0.0271)  (33.176,     -0.0921)  (33.319,     -0.1913)
(33.462,     -0.2651)  (33.605,     -0.3109)  (33.748,     -0.3281)  (33.891,     -0.3186)  (34.034,     -0.2860)  (34.177,     -0.2354)
(34.320,     -0.1726)  (34.463,     -0.1038)  (34.606,     -0.0347)  (34.749,      0.0293)  (34.892,      0.0840)  (35.035,      0.1262)
};
\end{axis}
\end{tikzpicture}
\caption{The integrand of $\mathcal{I}(s)$ where $s=0.99$, $\nu=\frac{17}{2}$, $n_\gamma=9$, $n_x=5$, $\lambda=3$, $R_1=2.0$, $\zeta_1=1.5$, $R_2=5.5$, and $\zeta_2=2$ (see row 2 of Table \ref{TABLESG}).
}
\label{NucAttrac}
\end{figure}

Originally, Gauss-Laguerre quadrature is used~\cite{Trivedi-Steinborn-27-670-83, Grotendorst-Steinborn-38-3875-88}, almost completely ignoring the possible effects of the oscillations.

The semi-infinite integrals can be transformed into an infinite series:
\begin{equation}
{\cal I}(s) = \sum_{n=0}^{+\infty} \, \int_{j_{\lambda,v}^{n}}^{j_{\lambda,v}^{n+1}} x^{n_x} \, \frac{\hat{k}_{\nu}[R_2 \gamma_{12}(s,x)]}{[\gamma_{12}(s,x)]^{n_{\gamma}}} \, j_{\lambda}(v\,x){\rm\,d}x,
\label{EQSUMAT}
\end{equation}
where $j_{\lambda,v}^{0} \,=\, 0$ and $j_{\lambda,v}^{n}=j_{\lambda+\frac{1}{2}}^{n}/v$ ($v \not= 0$) for $n=1,2,\ldots$ are the successive positive zeros of $j_{\lambda}(vx)$, where $j_{\lambda+\frac{1}{2}}^{n}$ for $n=1,2,\ldots$ are the successive positive zeros of $j_{\lambda}(x)$.

Unfortunately, these infinite series are slowly convergent. The use of Levin's $u$ transform~\cite{Levin-B3-371-73} or the epsilon algorithm of Wynn~\cite{Wynn-5-160-62}, which are considered among the most powerful convergence accelerators, could improve convergence of the infinite series. Unfortunately, the calculation times are still prohibitive for a high pre-determined accuracy. Then, the extrapolation methods are implemented~\cite{Safouhi9, Safouhi21, Safouhi31, Safouhi32, Safouhi35, Safouhi38}, which use numerical integration of successive intervals to construct approximations the semi-infinite integral. While these methods are generally highly accurate, there are some ranges of parameters where the computation becomes extremely heavy.

\section{The analytical development}
We begin by considering integrals of the form~\cite{Watson-44}:
\begin{equation}
\int_0^\infty x^{\eta+1} J_\eta(\beta\,x) \frac{K_\nu\left(\alpha\sqrt{x^2+z^2}\right)}{\sqrt{(x^2+z^2)^\nu}} {\rm\,d}x,
\label{EQMODEL1}
\end{equation}
where $\nu\in\mathbb{R}$, ${\rm Re}\,\eta>-1$, $\alpha>0$, $\beta>0$, and $|\arg z|<\frac{\pi}{2}$.

In these integrals, cylindrical Bessel functions are used instead of spherical Bessel functions. These integrals have an analytical expression in Watson's Treatise on the Theory of Bessel functions~\cite[\S 13.47]{Watson-44} that can be proved easily.

We begin by inserting the integral representation of the modified Bessel function~\eqref{EQINTREPMODBESS} into the expression~\eqref{EQMODEL1} and we reverse the order of integration:
\begin{align}
\int_0^\infty x^{\eta+1} J_\eta(\beta\,x) \frac{K_\nu\left(\alpha\sqrt{x^2+z^2}\right)}{\sqrt{(x^2+z^2)^\nu}} {\rm\,d}x
& = \dfrac{1}{2}\int_0^\infty \int_0^\infty x^{\eta+1} J_\eta(\beta\,x) \dfrac{e^{-\frac{\alpha}{2}(t+\frac{x^2+z^2}{t})}}{t^{\nu+1}}{\rm\,d}t {\rm\,d}x\\
& = \dfrac{1}{2}\int_0^\infty \dfrac{e^{-\frac{\alpha}{2}(t+\frac{z^2}{t})}}{t^{\nu+1}} \int_0^\infty x^{\eta+1}  e^{-\frac{\alpha x^2}{2t}}J_\eta(\beta\,x) {\rm\,d}x{\rm\,d}t,
\end{align}
By using the convergent series for the Bessel function, the inner integrals are all readily obtainable~\cite[\S 6.22]{Watson-44}:
\begin{align}
\int_0^\infty x^{\eta+1} e^{-\frac{\alpha x^2}{2t}}J_\eta(\beta\,x) {\rm\,d}x
& = \int_0^\infty e^{-\frac{\alpha x^2}{2t}}\sum_{k=0}^\infty \dfrac{(-1)^k}{k!\,\Gamma(k+\eta+1)}\left(\dfrac{\beta}{2}\right)^{2k+\eta}x^{2k+2\eta+1}{\rm\,d}x
\nonumber\\
& = \sum_{k=0}^\infty \dfrac{(-1)^k}{k!\,\Gamma(k+\eta+1)}\left(\dfrac{\beta}{2}\right)^{2k+\eta}\int_0^\infty e^{-\frac{\alpha x^2}{2t}}x^{2k+2\eta+1}{\rm\,d}x
\nonumber\\
& = \sum_{k=0}^\infty \dfrac{(-1)^k}{k!\,\Gamma(k+\eta+1)}\left(\dfrac{\beta}{2}\right)^{2k+\eta}\dfrac{\Gamma(k+\eta+1)}{2}\left(\dfrac{2t}{\alpha}\right)^{k+\eta+1}
\nonumber\\
& = \dfrac{\beta^\eta t^{\eta+1}}{\alpha^{\eta+1}}\sum_{k=0}^\infty \dfrac{(-1)^k}{k!}\left(\dfrac{\beta^2t}{2\alpha}\right)^{k}
\nonumber\\
& = \dfrac{\beta^\eta t^{\eta+1}}{\alpha^{\eta+1}}\exp\left(-\dfrac{\beta^2t}{2\alpha}\right).
\end{align}

Then, the outer integral is evaluated by recognizing it as a modified Bessel function, and we obtain:
\begin{align}
\int_0^\infty x^{\eta+1} J_\eta(\beta\,x)\frac{K_\nu\left(\alpha\sqrt{x^2+z^2}\right)}{\sqrt{(x^2+z^2)^\nu}} {\rm\,d}x
& = \dfrac{\beta^\eta}{2\alpha^{\eta+1}}\int_0^\infty \dfrac{e^{-\frac{\alpha}{2}(t+\frac{z^2}{t})-\frac{\beta^2t}{2\alpha}}}{t^{\nu-\eta}}{\rm\,d}t
\nonumber\\
& = \frac{\beta^\eta}{\alpha^\nu}\left(\frac{\sqrt{\alpha^2+\beta^2}}{z}\right)^{\nu-\eta-1}K_{\nu-\eta-1}\left(z\sqrt{\alpha^2+\beta^2}\right).
\label{eq:TCcyl}
\end{align}

\subsection{The semi-infinite integral ${\cal I}(s)$}

Let the integral ${\cal I}_{\lambda,r}^{\nu,\mu}(\alpha,\beta,z)$ be given by:
\begin{equation}
{\cal I}_{\lambda,r}^{\nu,\mu}(\alpha,\beta,z) = \int_0^\infty x^{\lambda+2r+2}\frac{\hat{k}_\nu\left[\alpha\sqrt{x^2+z^2}\right]}{(x^2+z^2)^{\nu-\mu}}j_{\lambda}(\beta\,x){\rm\,d}x.
\label{EQIlrnm}
\end{equation}

The simpler case of ${\cal I}_{\lambda,r}^{\nu,\mu}(\alpha,\beta,z)$ is given by:
\begin{equation}
{\cal I}_{\lambda,0}^{\nu,0}(\alpha,\beta,z) = \int_0^\infty x^{\lambda+2}\frac{\hat{k}_\nu\left[\alpha\sqrt{x^2+z^2}\right]}{(x^2+z^2)^{\nu}}j_{\lambda}(\beta\,x){\rm\,d}x.
\end{equation}

We can also write:
\begin{equation}\label{eq:TCNAI1-0}
{\cal I}_{\lambda+r,0}^{\nu-\mu,0}(\alpha,\beta,z) = \int_0^\infty x^{\lambda+r+2}\frac{\hat{k}_{\nu-\mu}\left[\alpha\sqrt{x^2+z^2}\right]}{(x^2+z^2)^{\nu-\mu}}j_{\lambda+r}(\beta\,x){\rm\,d}x,
\end{equation}

This simpler case is the essential starting point because identities can be used to increase the $r$ and $\mu$ parameters to obtain an expression for ${\cal I}_{\lambda,r}^{\nu,\mu}(\alpha,\beta,z)$. If $r, \mu \geq 0$, then by applying the following identities, which follow directly from equations~\eqref{eq:Krecrel} and \eqref{eq:Jrecrel}:
\begin{equation}
\frac{1}{\beta^{\lambda+1}}\left(\frac{\partial}{\beta\partial\beta}\right)^r\left(\beta^{\lambda+r+1}j_{\lambda+r}(\beta\,x)\right) = x^rj_\lambda(\beta\,x),
\end{equation}
and:
\begin{equation}
(-1)^\mu\alpha^{2\nu}\left(\frac{\partial}{\alpha\partial\alpha}\right)^\mu\left(\alpha^{2\mu-2\nu}\hat{k}_{\nu-\mu}\left[\alpha\sqrt{x^2+z^2}\right]\right) = \hat{k}_\nu\left[\alpha\sqrt{x^2+z^2}\right],
\end{equation}
to the integral ${\cal I}_{\lambda+r,0}^{\nu-\mu,0}(\alpha,\beta,z)$~\eqref{eq:TCNAI1-0}, we obtain:
\begin{equation}\label{eq:TCNAI1}
{\cal I}_{\lambda,r}^{\nu,\mu}(\alpha,\beta,z) = \frac{(-1)^\mu\alpha^{2\nu}}{\beta^{\lambda+1}} \left(\frac{\partial}{\alpha\partial\alpha}\right)^\mu \left(\frac{\partial}{\beta\partial\beta}\right)^r \left(\alpha^{2\mu-2\nu}\beta^{\lambda+r+1} {\cal I}_{\lambda+r,0}^{\nu-\mu,0}(\alpha,\beta,z)\right).
\end{equation}

Upon changing the Bessel functions to their spherical and reduced counterparts in~\eqref{eq:TCcyl}, we obtain:
\begin{eqnarray}
{\cal I}_{\lambda,0}^{\nu,0}(\alpha,\beta,z) & = & \int_0^\infty x^{\lambda+2}\frac{\hat{k}_\nu\left[\alpha\sqrt{x^2+z^2}\right]}{(x^2+z^2)^{\nu}}j_{\lambda}(\beta\,x){\rm\,d}x
\nonumber\\ & = & \int_0^\infty x^{\lambda+2} \frac{\sqrt{\frac{2}{\pi}} \left[\alpha\sqrt{x^2+z^2}\right]^{\nu} K_{\nu}\left[\alpha\sqrt{x^2+z^2}\right]}{(x^2+z^2)^{\nu}}
\sqrt{\frac{\pi}{2\,\beta\,x}} J_{\lambda+1/2}(\beta\,x)
{\rm\,d}x
\nonumber\\ & = & \frac{\alpha^{\nu}}{ \sqrt{\beta} } \int_0^\infty x^{\lambda+2-\frac{1}{2}}
\frac{K_{\nu} \left[ \alpha \sqrt{x^2+z^2} \right] } { \sqrt{(x^2+z^2)^{\nu}} }
 J_{\lambda+1/2}(\beta\,x)
{\rm\,d}x.
\end{eqnarray}

By using the analytic expression obtained in equation~\eqref{eq:TCcyl}, we obtain:
\begin{equation}
{\cal I}_{\lambda,0}^{\nu,0}(\alpha,\beta,z) = \dfrac{\beta^\lambda(\alpha^2+\beta^2)^{\frac{\nu-\lambda-3/2}{2}}}{z^{\nu-\lambda-3/2}}K_{\nu-\lambda-3/2}\left(z\sqrt{\alpha^2+\beta^2}\right),
\end{equation}
and:
\begin{equation}
{\cal I}_{\lambda+r,0}^{\nu-\mu,0}(\alpha,\beta,z) = \dfrac{\beta^{\lambda+r}(\alpha^2+\beta^2)^{\frac{\nu-\mu-\lambda-r-3/2}{2}}}{z^{\nu-\mu-\lambda-r-3/2}} K_{\nu-\mu-\lambda-r-3/2}\left(z\sqrt{\alpha^2+\beta^2}\right).
\end{equation}

Then, upon inserting the expression for ${\cal I}_{\lambda+r,0}^{\nu-\mu,0}(\alpha,\beta,z)$ into~\eqref{eq:TCNAI1},
given the reflection formula $K_{\tau}(x) = K_{-\tau}(x)$, we obtain:
\begin{equation}\label{eq:TCNAI2}
{\cal I}_{\lambda,r}^{\nu,\mu}(\alpha,\beta,z) = \dfrac{(-1)^\mu\alpha^{2\nu}z^\tau}{\beta^{\lambda+1}} \left(\frac{\partial}{\alpha\partial\alpha}\right)^\mu \left(\frac{\partial}{\beta\partial\beta}\right)^r\left(\alpha^{2\mu-2\nu}\beta^{2\lambda+2r+1} \dfrac{K_{\tau}(z\sqrt{\alpha^2+\beta^2})}{(\alpha^2+\beta^2)^{\tau/2}}\right),
\end{equation}
where:
\begin{equation}\label{EQTAU}
\tau = \lambda + r + \mu - \nu+3/2.
\end{equation}

Evidently, all that remains to do is to expand the derivations in~\eqref{eq:TCNAI2}, and the expression will be obtained. We start with the derivations with respect to $\beta$.

Using the identities:
\begin{align}
\left(\frac{\partial}{\beta\partial\beta}\right)^{r-\sigma}\left(\beta^{2\lambda+2r+1}\right) & = (-2)^{r-\sigma}(-\lambda-r-1/2)_{r-\sigma}\beta^{2\lambda+2r-2(r-\sigma)+1}
\nonumber\\
\left(\frac{\partial}{\beta\partial\beta}\right)^{\sigma}\left(\dfrac{K_{\tau}(z\sqrt{\alpha^2+\beta^2})}{(\alpha^2+\beta^2)^{\tau/2}}\right) & = (-z)^{\sigma}\dfrac{K_{\tau+\sigma}(z\sqrt{\alpha^2+\beta^2})}{(\alpha^2+\beta^2)^{(\tau+\sigma)/2}},
\end{align}
for some $\sigma\in\mathbb{N}_0$, the product rule yields:
\begin{align}
\left(\frac{\partial}{\beta\partial\beta}\right)^r &\left(\beta^{2\lambda+2r+1} \dfrac{K_{\tau}(z\sqrt{\alpha^2+\beta^2})}{(\alpha^2+\beta^2)^{\tau/2}}\right)
\nonumber\\
& = \sum_{\sigma=0}^r\binom{r}{\sigma} \left(\frac{\partial}{\beta\partial\beta}\right)^{r-\sigma} \left(\beta^{2\lambda+2r+1}\right)\left(\frac{\partial}{\beta\partial\beta}\right)^{\sigma} \left(\dfrac{K_{\tau}(z\sqrt{\alpha^2+\beta^2})}{(\alpha^2+\beta^2)^{\tau/2}}\right)
\nonumber\\
& = \sum_{\sigma=0}^r\binom{r}{\sigma}(-2)^{r-\sigma}(-\lambda-r-1/2)_{r-\sigma}\beta^{2\lambda+2r-2(r-\sigma)+1} \, (-z)^{\sigma} \dfrac{K_{\tau+\sigma}(z\sqrt{\alpha^2+\beta^2})}{(\alpha^2+\beta^2)^{(\tau+\sigma)/2}}
\nonumber\\
& = (-2)^{r} \beta^{2\lambda+1}
\sum_{\sigma=0}^r\binom{r}{\sigma} \left(\frac{\beta^{2} \, z}{2}\right)^{\sigma}(-\lambda-r-1/2)_{r-\sigma}
\dfrac{K_{\tau+\sigma}(z\sqrt{\alpha^2+\beta^2})}{(\alpha^2+\beta^2)^{(\tau+\sigma)/2}}.
\end{align}

We continue with the derivations with respect to $\alpha$. Using the identities:
\begin{align}
\left(\frac{\partial}{\alpha\partial\alpha}\right)^{\mu-m}\left(\alpha^{2\mu-2\nu}\right) & = (-2)^{\mu-m}(\nu-\mu)_{\mu-m}\alpha^{2\mu-2\nu-2(\mu-m)}
\nonumber\\
\left(\frac{\partial}{\alpha\partial\alpha}\right)^{m}\left(\dfrac{K_{\tau+s}(z\sqrt{\alpha^2+\beta^2})}{(\alpha^2+\beta^2)^{(\tau+s)/2}}\right) & = (-z)^{m}\dfrac{K_{\tau+s+m}(z\sqrt{\alpha^2+\beta^2})}{(\alpha^2+\beta^2)^{(\tau+s+m)/2}},
\end{align}
for some $m\in\mathbb{N}_0$, the product rule yields:
\begin{align}
\left(\frac{\partial}{\alpha\partial\alpha}\right)^\mu&\left(\alpha^{2\mu-2\nu} \dfrac{K_{\tau+s}(z\sqrt{\alpha^2+\beta^2})}{(\alpha^2+\beta^2)^{(\tau+s)/2}}\right)
\nonumber\\
& = \sum_{m=0}^\mu \binom{\mu}{m} \left(\frac{\partial}{\alpha\partial\alpha}\right)^{\mu-m} \left(\alpha^{2\mu-2\nu}\right)\left(\frac{\partial}{\alpha\partial\alpha}\right)^{m} \left(\dfrac{K_{\tau+s}(z\sqrt{\alpha^2+\beta^2})}{(\alpha^2+\beta^2)^{(\tau+s)/2}}\right)
\nonumber\\
& = \sum_{m=0}^\mu \binom{\mu}{m}(-2)^{\mu-m}(\nu-\mu)_{\mu-m}\alpha^{-2\nu+2m}(-z)^{m} \dfrac{K_{\tau+s+m}(z\sqrt{\alpha^2+\beta^2})}{(\alpha^2+\beta^2)^{(\tau+s+m)/2}}
\nonumber\\
& = \alpha^{-2\nu} \sum_{m=0}^\mu\binom{\mu}{m} \left(\dfrac{\alpha^2z}{2}\right)^{m}(\nu-\mu)_{\mu-m} \dfrac{K_{\tau+s+m}(z\sqrt{\alpha^2+\beta^2})}{(\alpha^2+\beta^2)^{(\tau+s+m)/2}}.
\end{align}

Then, combining these results and inserting them in~\eqref{eq:TCNAI2}, we obtain:
\begin{align}\label{eq:TCNAI3}
{\cal I}_{\lambda,r}^{\nu,\mu}(\alpha,\beta,z) & =
(-2)^r \alpha^{-2\nu} z^{\tau}\beta^{\lambda+1}\sum_{\sigma=0}^r \binom{r}{\sigma}\left(\dfrac{\beta^2z}{2}\right)^{\sigma}(-\lambda-r-1/2)_{r-\sigma}
\nonumber\\ &\times \sum_{m=0}^\mu \binom{\mu}{m} \left(\dfrac{\alpha^2z}{2}\right)^{m}(\nu-\mu)_{\mu-m} \dfrac{K_{\tau+\sigma+m}(z\sqrt{\alpha^2+\beta^2})}{(\alpha^2+\beta^2)^{(\tau+\sigma+m)/2}}.
\end{align}

Finally, replacing $\tau$ by its original value given by~\eqref{EQTAU}, we obtain:
\begin{align}\label{eq:TCNAI4}
{\cal I}_{\lambda,r}^{\nu,\mu}(\alpha,\beta,z) & = (-2)^r \alpha^{-2\nu} z^{\lambda+r+\mu-\nu+3/2} \beta^{\lambda+1} \sum_{\sigma=0}^r\binom{r}{\sigma} \left(\dfrac{\beta^2z}{2}\right)^{\sigma}
(-\lambda-r-1/2)_{r-\sigma}
\nonumber\\ &\times\sum_{m=0}^\mu \binom{\mu}{m} \left(\dfrac{\alpha^2z}{2}\right)^{m}(\nu-\mu)_{\nu-m}
\dfrac{K_{\lambda+r+\mu-\nu+\sigma+m+3/2}(z\sqrt{\alpha^2+\beta^2})}{(\alpha^2+\beta^2)^{(\lambda+r+\mu\nu+\sigma+m-+3/2)/2}}.
\end{align}

Now, the semi-infinite integrals ${\cal I}(s)$~\eqref{EQANATSIBESS} are formulated using the spherical and reduced Bessel functions, and they can be expressed as follows:
\begin{equation}\label{eq:TCNAI4INT1}
{\cal I}(s) = \frac{1}{\left[s\, (1-s)\right]^{\frac{n_{\gamma}}{2}}} \, {\cal I}_{\lambda,r}^{\nu, \mu}(R_2 \sqrt{s (1-s)},v,z),
\end{equation}
where $\nu$, $n{\gamma}$, $n_x$, $v$, $\lambda$ and $s$ are given by~\eqref{EQANATPARAM}. The integers $\mu$ and $r$, and the parameter $z$ are given by:
\begin{equation}
r = \dfrac{n_x-\lambda}{2}-1, \quad \mu=\dfrac{2\nu- n_{\gamma}}{2} \quad \textrm{and} \quad z = \frac{(1-s)\zeta_{1}^{2}+s\zeta_{2}^{2}}{s(1-s)}.
\end{equation}
Using the expressions of $\nu$, $n_{\gamma}$ and $n_x$  given by~\eqref{EQANATPARAM}, the summation indices $\lambda$, $l$, $l_1^{\prime}$, $l_2^{\prime}$ and $j$
involed in~\eqref{EQANAT} and defined by equation~\eqref{EQSUMINDEX}, one can easily verify that the integer $\mu \,\geq\, 0$.

However, the integer $r$  can be negative, but this only happens when $\lambda = n_x$ leading to $r = -1$. In this case, the analytical development is no longer valid and a different method should be used to compute the corresponding semi-infinite integral ${\cal I}(s)$.

Using equation~\eqref{eq:TCNAI4INT1} and if $\lambda \not= n_x $, we obtain:
\begin{align}\label{eq:TCNAI4INT2}
{\cal I}(s) & =  \frac{(-2)^{\frac{n_x-\lambda}{2}-1} \, \left(\sqrt{z}\right)^{n_x+\lambda-n_{\gamma}+1} v^{\lambda+1}}{R_2^{2\nu} \left[s\, (1-s)\right]^{\nu+\frac{n_{\gamma}}{2}}}
\nonumber\\ & \times  \sum_{\sigma=0}^{\frac{n_x-\lambda}{2}-1} \binom{\frac{n_x-\lambda}{2}-1}{\sigma} \left(\dfrac{v^2 \,z}{2}\right)^{\sigma}
\left(\frac{-n_x-\lambda+1}{2}\right)_{\frac{n_x-\lambda}{2}-1-\sigma} \;
\nonumber\\ & \times \sum_{m=0}^{\frac{2\nu- n_{\gamma}}{2}} \binom{\frac{2\nu- n_{\gamma}}{2}}{m} \left(\dfrac{R_2^2 \, s \, (1-s) \, z}{2}\right)^{m} \; (\nu-\mu)_{\nu-m}
\nonumber\\ & \hskip 3cm \times \dfrac{K_{\frac{n_x+\lambda-n_{\gamma}}{2}+\sigma+m+\frac{1}{2}}\left(z\sqrt{R_2^2 \, s  (1-s) +v^2}\right)}
{\left(\sqrt{R_2^2 \,s (1-s)+v^2}\right)^{\frac{n_x+\lambda-n_{\gamma}}{2}+\sigma+m+\frac{1}{2}}}.
\end{align}

For the case $r=-1$ corresponding to $n_x = \lambda$ in~\eqref{EQANATSIBESS}, one could use a most recent approach introduced
in~\cite{Safouhi55} and based on the $S$ transformation and the double exponential transformation. Other methods based on extrapolation methods and nonlinear transformations could also be used~\cite{Safouhi9, Safouhi38}.

\section{Numerical Discussion}
Since the order of the Bessel function is always an integer in~\eqref{eq:TCNAI4INT2}, the accuracy and efficiency of this formula are directly related to that of the algorithm for the calculation of a sequence of Bessel functions $\{K_\ell(\cdot)\}_{\ell\in\mathbb{N}_0}$.

Given the recurrence relation~\eqref{eq:Krecrel} is stable in the upward direction, and given the reflection formula $K_{\nu}(z) = K_{-\nu}(z)$ an efficient algorithm would provide the seed values $K_0(z)$ and $K_1(z)$ and compute the remaining values by recurrence. In~\cite[\S 6.5]{Press-07}, a very fast algorithm for these seed values is derived based on the use of rational minimax approximants. These approximants are rational functions with unknowns that are designed to minimize the maximum error on an interval. The algorithm achieves double precision accuracy by expressing $K_0(z)$ and $K_1(z)$ for $z\le 1$ as:
\begin{equation}
K_0(z) \approx \dfrac{p_4^{(0)}(z^2)}{q_2^{(0)}(1-z^2)} - \dfrac{\log(z) \,r_4^{(0)}(z^2)}{s_2^{(0)}(1-z^2)}
\quad \textrm{and} \quad
K_1(z) \approx \dfrac{z\, p_4^{(1)}(z^2)}{q_2^{(1)}(1-z^2)} + \dfrac{z\, \log(z) \, r_4^{(1)}(z^2)}{s_2^{(1)}(1-z^2)}+\dfrac{1}{z},
\end{equation}
and for $z>1$ as:
\begin{equation}
K_0(z) \approx \dfrac{\exp(-z)}{\sqrt{z}}\dfrac{u_7^{(0)}(z^{-1})}{v_7^{(0)}(z^{-1})}
\quad \textrm{and} \quad
K_1(z) \approx \dfrac{\exp(-z)}{\sqrt{z}}\dfrac{u_7^{(1)}(z^{-1})}{v_7^{(1)}(z^{-1})},
\end{equation}
where $p_n^{(m)}(z)$ stands for a polynomial of degree $n$ in $z$ and where $m=0,1$ serves to distinguish between the polynomial for $K_0(z)$ and that for $K_1(z)$. The coefficients of these approximants are given in~\cite{Num-Rec-Webnote-Bessik-07}.

Tables \ref{TABLESP} and \ref{TABLESG} contain values obtained for the semi-infinite integrals ${\cal I}(s)$~\eqref{EQANATSIBESS}. The calculations for values $s=0.05$ which is close to $0$, and $s=0.99$ which is close to $1$.

In Tables~\ref{TABLESP} and \ref{TABLESG}, we listed values of the semi-infinite integrals ${\cal I}(s)$ obtained by a general MATLAB built-in numerical integration function that uses global adaptive quadrature, set to an accuracy of $15$ correct digits. However, the MATLAB built-in function is not always able to complete the integral approximation, particularly for large values of $\lambda$ and/or $v$. In contrast, the analytical expression is able to complete the computation to machine accuracy regardless of the values of the parameters.

To further illustrate the high efficiency of the analytic expressions, we computed the semi-infinite integrals ${\cal I}(s)$ using two existing methods proven to be accurate and efficient, namely the combination of $S$ (See~\cite{Safouhi9}) and $\bar{D}$ transformations~\cite{Sidi-38-299-82}, and the transformation $\bar{D}$ with the $W$ algorithm of Sidi~\cite{Sidi-78-125-97}.

As can be seen from the numerical tables, the analytic expressions, which achieve machine accuracy, lead to a substantial gain in the calculation time. The gain in efficiency is of the order of $800$.

\section{Conclusion}

In this work, we report the analytical expression~\eqref{eq:TCNAI4} for the semi-infinite integral bottleneck occurring in the three-center nuclear attraction integrals over $B$ functions. We describe how to compute the formula to obtain an efficient evaluation in double precision arithmetic. This requires the rational minimax approximants that minimize the maximum error on the interval of evaluation. The numerical tests show a substantial gain in efficiency over existing methods based on nonlinear transformations and extrapolation methods.

\section{Numerical tables}

\begin{table}[!h]
\begin{center}
\caption{Evaluation of ${\cal I}(s)$~\eqref{EQANATSIBESS} involved in~(\ref{EQANAT}). $s = 0.05$, $\zeta_1=1.5$, $\zeta_2=2.0$, $R_1=2.0$, $R_2=5.5$ and $v=3.225$\label{TABLESP}}
\begin{tabular*}{\hsize}{@{\extracolsep{\fill}}cccccc} \sphline 
$ \nu-\frac{1}{2} $ & $n_{\gamma}$ & $n_x$ & $\lambda$ & ${\cal I}(s)^{\dag}$ & ${\cal I}(s)^{\textrm{Matlab}}$\\ \sphline
 7 & 11 &  3 &  1 & ~0.151~189~722~612~165~(-1) & ~0.151~189~722~613~836~(-1) \\
 7 & 11 &  4 &  0 & -0.770~700~245~226~897~(-1) & -0.770~700~245~226~550~(-1) \\
 7 & 11 &  4 &  2 & ~0.911~341~847~656~817~(-1) & ~0.911~341~847~657~070~(-1) \\
 6 & 13 &  3 &  1 & ~0.862~532~316~505~739~(-3) & ~0.862~532~316~513~426~(-3) \\
 6 & 13 &  4 &  0 & -0.412~000~772~378~986~(-2) & -0.412~000~772~377~796~(-2) \\
 6 & 13 &  4 &  2 & ~0.492~236~336~705~101~(-2) & ~0.492~236~336~704~074~(-2) \\
 7 & 15 &  3 &  1 & ~0.106~814~986~690~961~(-1) & ~0.106~814~986~691~068~(-1) \\
 9 & 17 &  3 &  1 & ~0.266~323~983~838~913~(~1) & ~0.266~323~983~839~022~(~1) \\
 9 & 19 &  3 &  1 & ~0.184~547~358~116~701~(~1) & ~0.184~547~358~116~765~(~1) \\
 9 & 19 &  4 &  0 & -0.762~928~846~920~085~(~1) & -0.762~928~846~919~988~(~1) \\
 9 & 19 &  5 &  1 & -0.347~485~191~071~318~(~2) & -0.347~485~191~071~172~(~2) \\
10 & 21 &  3 &  1 & ~0.257~290~058~890~616~(~2) & ~0.257~290~058~890~630~(~2) \\
10 & 21 &  4 &  0 & -0.101~363~984~175~823~(~3) & -0.101~363~984~175~821~(~3) \\
10 & 21 &  4 &  2 & ~0.125~297~943~142~392~(~3) & ~0.125~297~943~142~393~(~3) \\
10 & 21 &  5 &  1 & -0.440~284~382~122~921~(~3) & -0.440~284~382~122~792~(~3) \\
10 & 19 &  4 &  0 & -0.163~020~781~236~490~(~3) & -0.163~020~781~236~467~(~3) \\
10 & 19 &  5 &  1 & -0.746~484~060~054~242~(~3) & -0.746~484~060~053~884~(~3) \\
11 & 23 &  3 &  1 & ~0.372~866~539~760~214~(~3) & ~0.372~866~539~760~223~(~3) \\
11 & 21 &  4 &  0 & -0.235~942~964~388~561~(~4) & -0.235~942~964~388~555~(~4) \\
11 & 23 &  4 &  2 & ~0.174~686~408~404~681~(~4) & ~0.174~686~408~404~684~(~4) \\
11 & 23 &  5 &  1 & -0.579~875~446~713~486~(~4) & -0.579~875~446~713~279~(~4) \\
11 & 23 &  5 &  3 & ~0.850~707~087~650~976~(~4) & ~0.850~707~087~650~815~(~4) \\ \hline
\end{tabular*}
\begin{tablenotes}
\begin{itemize}
\item $\tilde{\cal I}(s)^{\dag}$ are obtained using the analytic expressions. $\mathcal{I}(s)^{\textrm{Matlab}}$ are obtained by means of a MATLAB built-in numerical integration function that uses global adaptive quadrature. Numbers in parentheses represent powers of $10$.
\item \underline{\bf Calculation times}: Comparisions with the combination $S-\bar{D}$ and the $W$ algorithm with the $\bar{D}$ transformation, illustrate the clear advantage of analytic expressions over extrapolation methods:
$$
\frac{\textrm{Total time using} \quad S\bar{D}}{\textrm{Total time using the analytic formulae}} \,=\,  779
\qquad \textrm{and} \qquad
\frac{\textrm{Total time using} \quad W\bar{D}}{\textrm{Total time using the analytic formulae}} \,=\, 722.
$$
\end{itemize}
\end{tablenotes}
\end{center}
\end{table}

\clearpage
\begin{table}
\begin{center}
\caption{Evaluation of ${\cal I}(s)$~\eqref{EQANATSIBESS} involved in~(\ref{EQANAT}). $s = 0.99$, $\zeta_1=1.5$, $\zeta_2=2.0$, $R_1=2.0$, $R_2=5.5$ and $v=1.9450$\label{TABLESG}}
\begin{tabular*}{\hsize}{@{\extracolsep{\fill}}cccccc} \sphline 
$ \nu-\frac{1}{2} $ & $n_{\gamma}$ & $n_x$ & $\lambda$ & ${\cal I}(s)^{\dag}$ & ${\cal I}(s)^{\textrm{Matlab}}$\\ \sphline
 8 &  9 &  5 &  1 & -0.136~578~999~110~210~(-3) & -0.136~578~513~775~021~(-3) \\
 8 &  9 &  5 &  3 & ~0.157~429~717~614~749~(-3) & ~0.157~428~199~599~963~(-3) \\
 9 & 11 &  4 &  0 & -0.133~767~585~018~979~(-3) & -0.133~767~051~011~091~(-3) \\
 9 & 11 &  5 &  1 & -0.238~345~682~495~741~(-2) & -0.238~345~138~495~788~(-2) \\
 9 & 11 &  5 &  3 & ~0.275~872~683~252~550~(-2) & ~0.275~872~094~789~520~(-2) \\
 9 & 11 &  4 &  2 & ~0.145~980~032~943~987~(-3) & ~0.145~979~955~050~279~(-3) \\
 9 & 13 &  5 &  1 & -0.189~782~159~585~373~(-2) & -0.189~782~160~850~882~(-2) \\
 9 & 17 &  3 &  1 & ~0.331~772~864~261~456~(-5) & ~0.331~772~945~380~560~(-5) \\
 9 & 19 &  3 &  1 & ~0.201~688~439~270~122~(-5) & ~0.201~688~456~114~013~(-5) \\
 9 & 19 &  4 &  0 & -0.301~438~469~081~214~(-4) & -0.301~438~463~736~758~(-4) \\
 9 & 19 &  5 &  1 & -0.476~213~479~931~115~(-3) & -0.476~213~471~529~263~(-3) \\
10 & 21 &  3 &  1 & ~0.246~169~752~226~837~(-4) & ~0.246~169~761~624~060~(-4) \\
10 & 21 &  4 &  0 & -0.358~276~851~579~279~(-3) & -0.358~276~848~288~597~(-3) \\
10 & 21 &  4 &  2 & ~0.396~246~479~172~108~(-3) & ~0.396~246~476~283~224~(-3) \\
10 & 21 &  5 &  1 & -0.551~474~883~342~359~(-2) & -0.551~474~878~125~191~(-2) \\
10 & 19 &  4 &  0 & -0.651~321~577~489~068~(-3) & -0.651~321~566~692~786~(-3) \\
10 & 19 &  5 &  1 & -0.103~223~435~438~069~(-1) & -0.103~223~433~320~636~(-1) \\
11 & 23 &  3 &  1 & ~0.307~741~810~026~843~(-3) & ~0.307~741~813~898~588~(-3) \\
11 & 21 &  4 &  0 & -0.843~566~525~213~554~(-2) & -0.843~566~520~507~011~(-2) \\
11 & 23 &  4 &  2 & ~0.483~636~533~415~352~(-2) & ~0.483~636~533~188~303~(-2) \\
11 & 23 &  5 &  1 & -0.654~156~086~743~768~(-1) & -0.654~156~083~528~505~(-1) \\
11 & 23 &  5 &  3 & ~0.778~484~244~434~089~(-1) & ~0.778~484~240~745~456~(-1) \\ \hline
\end{tabular*}
\begin{tablenotes}
\begin{itemize}
  \item $\tilde{\cal I}(s)^{\dag}$ are obtained using the analytic expressions. $\mathcal{I}(s)^{\textrm{Matlab}}$ are obtained by means of a MATLAB built-in numerical integration function that uses global adaptive quadrature. Numbers in parentheses represent powers of $10$.
  \item \underline{\bf Calculation times}: Comparisions with the combination $S-\bar{D}$ and the $W$ algorithm with the $\bar{D}$ transformation, illustrate the clear advantage of analytic expressions over extrapolation methods:
$$
\frac{\textrm{Total time using} \quad S\bar{D}}{\textrm{Total time using the analytic formulae}} \,=\,  848
\qquad \textrm{and} \qquad
\frac{\textrm{Total time using} \quad W\bar{D}}{\textrm{Total time using the analytic formulae}} \,=\, 766.
$$
\end{itemize}
\end{tablenotes}
\end{center}
\end{table}

\clearpage

\begin{thebibliography}{10}

\bibitem{Sack-5-252-64}
R.A. Sack.
\newblock Three-dimensional addition theorem for arbitrary functions involving
  expansions in spherical harmonics.
\newblock {\em J. Math. Phys.}, 5:252--259, 1964.

\bibitem{Sack-5-260-64}
R.A. Sack.
\newblock Two-center expansion for the powers of the distance between two
  points.
\newblock {\em J. Math. Phys.}, 5:260--268, 1964.

\bibitem{Sack-8-1774-67}
R.A. Sack.
\newblock Expansions in spherical harmonics. {IV}. integral form of the radial
  dependence.
\newblock {\em J. Math. Phys.}, 8:1774--1784, 1967.

\bibitem{Sack-5-774-74}
R.A. Sack.
\newblock Generating functions for spherical harmonics. {Part I}:
  Three-dimensional harmonics.
\newblock {\em SIAM J. Math. Anal.}, 5:774--796, 1974.

\bibitem{Hellmann-1937}
H.~Hellmann.
\newblock {\em Einf\"{u}hrung in die Quantenchemie}.
\newblock Deuticke, Leipzig, 1937.

\bibitem{Boys-200-542-50}
S.F. Boys.
\newblock Electronic wave functions. {I}. a general method of calculation for
  the stationary states of any molecular system.
\newblock {\em Proc. R. Soc. Lond. Series A, Math. \& Phys. Sciences.},
  200:542--554, 1950.

\bibitem{Boys-201-125-50}
S.F. Boys.
\newblock Electronic wave functions. {II}. a calculation for the ground state
  of the {Beryllium} atom.
\newblock {\em Proc. R. Soc. Lond. Series A, Math. \& Phys. Sciences.},
  201:125--137, 1950.

\bibitem{Agmon-85}
S.~Agmon.
\newblock {\em Bounds on exponential decay of eigenfunctions of {Schr\"odinger}
  operators}.
\newblock in S. Graffi (editor), {Schr\"odinger} operators. Springer-Verlag,
  Berlin, 1985.

\bibitem{Kato-10-151-57}
T.~Kato.
\newblock On the eigenfunctions of many-particle systems in quantum mechanics.
\newblock {\em Commun. Pure Appl. Math.}, 10:151--177, 1957.

\bibitem{Slater-42-33-32}
J.C. Slater.
\newblock Analytic atomic wave functions.
\newblock {\em Phys. Rev.}, 42:33--43, 1932.

\bibitem{Shavitt-63}
I.~Shavitt.
\newblock {\em The Gaussian function in calculation of statistical mechanics
  and quantum mechanics, Methods in Computational Physics. 2. Quantum
  Mechanics}.
\newblock edited by B. Alder, S. Fernbach, M. Rotenberg, Academic Press, New
  York, 1963.

\bibitem{Steinborn-Filter-38-273-75}
E.O. Steinborn and E.~Filter.
\newblock Translations of fields represented by spherical-harmonics expansions
  for molecular calculations. {III}. {Translations} of reduced {Bessel}
  functions, {Slater}-type s-orbitals, and other functions.
\newblock {\em Theor. Chim. Acta.}, 38:273--281, 1975.

\bibitem{Filter-Steinborn-18-1-78}
E.~Filter and E.O. Steinborn.
\newblock Extremely compact formulas for molecular one-electron integrals and
  {Coulomb} integrals over {Slater}-type orbitals.
\newblock {\em Phys. Rev. A.}, 18:1--11, 1978.

\bibitem{Ozdogan-Ruiz-08}
{\em Recent Advances in Computational Chemistry. Molecular Integrals over
  {Slater} Orbitals}.
\newblock Editors: Telhat Ozdogan and Maria Belen Ruiz. Published by Transworld
  Research Network., Kerala, India, 2008.

\bibitem{Niehaus-Lopez-Rico-41-485205-08}
T.A. Niehaus, R.~L\'{o}pez, and J.F. Rico.
\newblock Efficient evaluation of the {Fourier} transform over products of
  {Slater}-type orbitals on different centers.
\newblock {\em J. Phys. A: Math. Theor.}, 41:485205--485219, 2008.

\bibitem{Fernandez-Ema-Lopez-Ramirez-78-83-00}
J.~Fern\'{a}ndez Rico, J.J. Fern\'{a}ndez, I.~Ema, R.~L\'{o}pez, and
  G.~Ram\'{i}rez.
\newblock Master formulas for two- amd three-center one electron integrals
  involving cartesian {GTO, STO, and BTO}.
\newblock {\em Int. J. Quantum Chem.}, 78:83--93, 1999.

\bibitem{Rico-Fernandez-Ema-Lopez-Ramirez-81-16-01}
J.F. Rico, J.J. Fern\'andez, I.~Ema, R.~L\'opez, and G.~Ram\'irez.
\newblock Four-center integrals for gaussian and exponential functions.
\newblock {\em Int. J. Quant. Chem.}, 81:16--18, 2001.

\bibitem{Fernandez-Aguado-Lopez-Ema-Ramirez-8-148-01}
J.J. Fern\'andez, R.~L\'opez, A.~Aguado, I.~Ema, and G.~Ram\'irez.
\newblock {SMILES} {Slater} molecular integrals for large electronic systems:
  New program for molecular calculations with {Slater} type orbitals.
\newblock {\em Int. J. Quant. Chem.}, 81:148--153, 2001.

\bibitem{Steinborn-Homeier-Ema-Lopez-Ramirez-76-244-00}
E.O. Steinborn, H.H.H. Homeier, I.~Ema, R.~L\'{o}pez, and G.~Ram\'{i}rez.
\newblock Molecular calculations with ${B}$ functions.
\newblock {\em Int. J. Quantum Chem.}, 76:244--251, 2000.

\bibitem{Steinborn-Weniger-11-509-77}
E.O. Steinborn and E.J. Weniger.
\newblock Advantages of reduced {Bessel} functions as atomic orbitals: An
  application to {H}$^+_2$.
\newblock {\em Int. J. Quantum Chem. Symp.}, 11:509--516, 1977.

\bibitem{Steinborn-Weniger-12-103-78}
E.O. Steinborn and E.J. Weniger.
\newblock Reduced {Bessel} functions as atomic orbitals: Some mathematical
  aspects and an {LCAO-MO} treatment of {HeH}$^{++}$.
\newblock {\em Int. J. Quantum Chem. Symp.}, 12:103--108, 1978.

\bibitem{Weniger-Steinborn-78-6121-83}
E.J. Weniger and E.O. Steinborn.
\newblock The {Fourier} transforms of some exponential-type functions and their
  relevance to multicenter problems.
\newblock {\em J. Chem. Phys.}, 78:6121--6132, 1983.

\bibitem{Weniger-Grotendorst-Steinborn-33-3688-86}
E.J. Weniger, J.~Grotendorst, and E.O. Steinborn.
\newblock Unified analytical treatement of overlap, two-center nuclear
  attraction and {Coulomb} integrals of ${B}$ functions via the
  {Fourier}-transform method.
\newblock {\em Phys. Rev. A.}, 33:3688--3705, 1986.

\bibitem{Grotendorst-Weniger-Steinborn-33-3706-86}
J.~Grotendorst, E.J. Weniger, and E.O. Steinborn.
\newblock Efficient evaluation of infinite-series representations for overlap,
  two-center nuclear attraction, and {Coulomb} integrals using nonlinear
  convergence accelerators.
\newblock {\em Phys. Rev. A.}, 33:3706--3726, 1986.

\bibitem{Weniger-Steinborn-73-323-88}
E.J. Weniger and E.O. Steinborn.
\newblock Overlap integrals of ${B}$ functions. a numerical study of infinite
  series representations and integrals representation.
\newblock {\em Theor. Chim. Acta.}, 73:323--336, 1988.

\bibitem{Safouhi9}
H.~Safouhi.
\newblock The properties of sine, spherical {Bessel} and reduced {Bessel}
  functions for improving convergence of semi-infinite very oscillatory
  integrals: The evaluation of three-center nuclear attraction integrals over
  ${B}$ functions.
\newblock {\em J. Phys. A: Math. Gen.}, 34:2801--2818, 2001.

\bibitem{Safouhi21}
H.~Safouhi.
\newblock Highly accurate numerical results for three-center nuclear attraction
  and two-electron {Coulomb} and exchange integrals over {Slater} type
  functions.
\newblock {\em Int. J. Quantum Chem.}, 100:172--183, 2004.

\bibitem{Safouhi31}
H.~Safouhi and A.~Bouferguene.
\newblock Nonlinear transformation methods for accelerating the convergence of
  {Coulomb} integrals over exponential type functions.
\newblock {\em Theo. Chem. Acc.}, 117:213--222, 2007.

\bibitem{Safouhi32}
S.~Duret and H.~Safouhi.
\newblock The $w$ algorithm and the $\bar{D}$ transformation for the numerical
  evaluation of three-center nuclear attraction integrals.
\newblock {\em Int. J. Quantum Chem.}, 107:1060--1066, 2007.

\bibitem{Safouhi35}
R.M. Slevinsky and H.~Safouhi.
\newblock The {$S$} and {$G$} transformations for computing three-center
  nuclear attraction integrals.
\newblock {\em Int. J. Quantum Chem.}, 109:1741--1747, 2009.

\bibitem{Safouhi38}
H.~Safouhi.
\newblock {Bessel}, sine and cosine functions and extrapolation methods for
  computing molecular multi-center integrals.
\newblock {\em Numer. Algor.}, 54:141--167, 2010.

\bibitem{Safouhi45}
R.M. Slevinsky and H.~Safouhi.
\newblock A comparative study of numerical steepest descent, extrapolation, and
  sequence transformation methods in computing semi-infinite integrals.
\newblock {\em Numer. Algor.}, 60:315--337, 2012.

\bibitem{Iserles-Norsett-44-755-04}
A.~Iserles and S.P. N{\o}rsett.
\newblock On quadrature methods for highly oscillatory integrals and their
  implementation.
\newblock {\em BIT}, 44:755--772, 2004.

\bibitem{Iserles-Norsett-461-1383-05}
A.~Iserles and S.P. N{\o}rsett.
\newblock Efficient quadrature of highly oscillatory integrals using
  derivatives.
\newblock {\em Proc. R. Soc. Lond. A}, 461:1383--1399, 2005.

\bibitem{Olver-26-213-06}
S.~Olver.
\newblock Moment-free numerical integration of highly oscillatory functions.
\newblock {\em IMA J. Num. Anal.}, 26:213--227, 2006.

\bibitem{Huybrechs-Vandewalle-44-1026-06}
D.~Huybrechs and S.~Vandewalle.
\newblock On the evaluation of highly oscillatory integrals by analytic
  continuation.
\newblock {\em SIAM J. Numer. Anal.}, 44:1026--1048, 2006.

\bibitem{Cools-Huybrechs-Nuyens-109-1748-09}
R.~Cools, D.~Huybrechs, and D.~Nuyens.
\newblock Recent topics in numerical integration.
\newblock {\em Int. J. Quant. Chem.}, 109:1748--1755, 2009.

\bibitem{Huybrechs-Olver-09}
D.~Huybrechs and S.~Olver.
\newblock {\em Highly Oscillatory Quadrature, in Highly Oscillatory Problems:
  Computation, Theory and Applications}.
\newblock Cambridge University Press, Cambridge, 2009.

\bibitem{Levin-Sidi-9-175-81}
D.~Levin and A.~Sidi.
\newblock Two new classes of non-linear transformations for accelerating the
  convergence of infinite integrals and series.
\newblock {\em Appl. Math. Comput.}, 9:175--215, 1981.

\bibitem{Gray-Atchison-4-363-67}
H.L. Gray and T.A. Atchison.
\newblock Nonlinear transformation related to the evaluation of improper
  integrals. {I}.
\newblock {\em SIAM J. Numer. Anal.}, 4:363--371, 1967.

\bibitem{Brezinski-Zaglia-91}
C.~Brezinski and M.~Redivo-Zaglia.
\newblock {\em Extrapolation Methods: Theory and Practice}.
\newblock Edition North-Holland, Amsterdam, 1991.

\bibitem{Gray-Wang-29-271-92}
H.L. Gray and S.~Wang.
\newblock A new method for approximating improper integrals.
\newblock {\em SIAM J. Numer. Anal.}, 29:271--283, 1992.

\bibitem{Sidi-03}
A.~Sidi.
\newblock {\em Practical Extrapolation Methods: Theory and Applications}.
\newblock Cambridge U. P., Cambridge, 2003.

\bibitem{Safouhi39}
R.M. Slevinsky, T.~Temga, M.~Mouattamid, and H.~Safouhi.
\newblock One- and two-center {ETF}-integrals of first order in relativistic
  calculation of {NMR} parameters.
\newblock {\em J. Phys. A: Math. Theor.}, 43:225202, 2010.

\bibitem{Safouhi33}
L.~Berlu and H.~Safouhi.
\newblock Analytical treatment of nuclear magnetic shielding tensor integrals
  over exponential type functions.
\newblock {\em J. Theor. Comp. Chem.}, 7:1215--1225, 2008.

\bibitem{Safouhi41}
H.~Safouhi.
\newblock Integrals of the paramagnetic contribution in the relativistic
  calculation of the shielding tensor.
\newblock {\em J. Math. Chem.}, 48:601--616, 2010.

\bibitem{Condon-Shortley-51}
E.U. Condon and G.H. Shortley.
\newblock {\em The theory of atomic spectra}.
\newblock Cambridge University Press, Cambridge, England, 1951.

\bibitem{Abramowitz-Stegun-65}
M.~Abramowitz and I.A. Stegun.
\newblock {\em Handbook of Mathematical Functions}.
\newblock Dover, New York, 1965.

\bibitem{Watson-44}
G.N. Watson.
\newblock {\em A Treatise on the Theory of {Bessel} Functions}.
\newblock Cambridge University Press, Second Edition, Cambridge, England, 1944.

\bibitem{Arfken-Weber-95}
G.B. Arfken and H.J. Weber.
\newblock {\em Mathematical methods for Physicists}.
\newblock Academic Press, Fifth edition, London, 1995.

\bibitem{Weissbluth-78}
M.~Weissbluth.
\newblock {\em Atoms and molecules}.
\newblock Academic, New York, 1978.

\bibitem{Gelfand-Shilov-64}
I.M. Gel'fand and G.E. Shilov.
\newblock {\em Generalized functions {I}, properties and operations}.
\newblock Academic, New York, 1964.

\bibitem{Gaunt-228-151-29}
J.A. Gaunt.
\newblock The triplets of helium.
\newblock {\em Phil. Trans. Roy. Soc.}, A. 228:151--196, 1929.

\bibitem{Weniger-Steinborn-25-149-82}
E.J. Weniger and E.O. Steinborn.
\newblock Programs for the coupling of spherical harmonics.
\newblock {\em Comput. Phys.Commun.}, 25:149--157, 1982.

\bibitem{Xu-65-1601-96}
Yu-Lin Xu.
\newblock Fast evaluation of {Gaunt} coefficients.
\newblock {\em Math. Comput.}, 65:1601--1612, 1996.

\bibitem{Xu-85-53-97}
Yu-Lin Xu.
\newblock Fast evaluation of {Gaunt} coefficients: recursive approach.
\newblock {\em J. Comput. Appl. Math.}, 85:53--65, 1997.

\bibitem{Trivedi-Steinborn-27-670-83}
H.P. Trivedi and E.O. Steinborn.
\newblock {Fourier} transform of a two-center product of exponential-type
  orbitals. application to one- and two-electron multicenter integrals.
\newblock {\em Phys. Rev. A.}, 27:670--679, 1983.

\bibitem{Grotendorst-Steinborn-38-3875-88}
J.~Grotendorst and E.O. Steinborn.
\newblock Numerical evaluation of molecular one- and two-electron multicenter
  integrals with exponential-type orbitals via the {Fourier}-transform method.
\newblock {\em Phys. Rev. A.}, 38:3857--3876, 1988.

\bibitem{Levin-B3-371-73}
D.~Levin.
\newblock Developement of non-linear transformations for improving convergence
  of sequences.
\newblock {\em Int. J. Comput. Math.}, B3:371--388, 1973.

\bibitem{Wynn-5-160-62}
P.~Wynn.
\newblock Upon a second confluent form the $\epsilon$-algorithm.
\newblock {\em Proc. Glascow Math. Assoc.}, 5:160--165, 1962.

\bibitem{Safouhi55}
J.~Lovrod and H.~Safouhi.
\newblock Double exponential transformation for computing three-center nuclear
  attraction integrals.
\newblock {\em Molecular Physics}, 118:1--12, 2020.

\bibitem{Press-07}
W.H. Press, S.A. Teukolsky, W.T. Vetterling, and B.P. Flannery.
\newblock {\em Numerical Recipes: The Art of Scientific Computing}.
\newblock Cambridge University Press, Third Edition, New York, 2007.

\bibitem{Num-Rec-Webnote-Bessik-07}
Numerical~Recipes Software.
\newblock {\em Coefficients Used in the Bessjy and Bessik Objects}.
\newblock Numerical Recipes Webnote No.~7, at http://www.nr.com/webnotes?7,
  2007.

\bibitem{Sidi-38-299-82}
A.~Sidi.
\newblock An algorithm for a special case of a generalization of the
  {Richardson} extrapolation process.
\newblock {\em Numer. Math.}, 38:299--307, 1982.

\bibitem{Sidi-78-125-97}
A.~Sidi.
\newblock Computation of infinite integrals involving {Bessel} functions of
  arbitrary order by the $\bar{D}$-transformation.
\newblock {\em J. Comp. Appl. Math.}, 78:125--130, 1997.

\end{thebibliography}

\end{document}